\documentclass{aa}
\usepackage{graphics}
\begin{document}


\title{The Extended X-ray Halo of the Crab-like SNR G21.5-0.9}
\author{R.S. Warwick\inst{1} 
\and J-P. Bernard\inst{2}
\and F. Bocchino\inst{3}
\and A. Decourchelle\inst{4}
\and P. Ferrando\inst{4}
\and R.G. Griffiths\inst{1}
\and F. Haberl\inst{5}
\and N. La Palombara\inst{6}
\and D. Lumb\inst{3}
\and S. Mereghetti\inst{6}
\and A.M.  Read\inst{5}
\and D. Schaudel\inst{5}
\and N. Schurch\inst{1}
\and A. Tiengo\inst{7}
\and R. Willingale\inst{1}
}

\offprints{R.S. Warwick, \email{rsw@star.le.ac.uk}}

\institute{Department of Physics and Astronomy,
University of Leicester, Leicester LE1 7RH, U.K.
\and Institute d'Astrophysique Spatiale, Orsay, France.
\and Space Science Department,  ESTEC, 2200 AG Noordwijk, 
The Netherlands.
\and Service d'Astrophysique, CEA Saclay, Gif-sur-Yvette, 91191, France.
\and Max-Planck-Institut f{\"u}r extraterrestrische Physik, D-85740,
Garching, Germany.
\and Istituto di Fisica Cosmica ``G.Occhialini'', CNR, I-20133, Milano, Italy.
\and  XMM-SOC, Villafranca Satellite Tracking Station, 28080 Madrid, Spain.
}

  \date{Received; Accepted}

  \titlerunning{{\em XMM-Newton} Observations of G21.5-0.9}

  \authorrunning{Warwick et al.}

\abstract{
Recent {\em XMM-Newton} observations reveal an extended 
($r \approx 150''$) 
low-surface brightness X-ray halo in the supernova remnant G21.5-0.9. 
The near circular symmetry, the lack of any limb brightening and the 
non-thermal spectral form, all favour an interpretation of this outer halo 
as an extension of the central synchrotron nebula rather than as a shell 
formed by the supernova blast wave and ejecta. The X-ray spectrum of
the nebula exhibits a marked spectral softening with radius, with
the power-law spectral index varying from $\Gamma = 1.63 \pm 0.04$ in the 
core to $\Gamma = 2.45 \pm 0.06$  at the edge of the halo. Similar spectral 
trends are seen in other Crab-like remnants and reflect the impact of 
the synchrotron radiation losses on very high energy electrons as they 
diffuse out from the inner nebula. A preliminary timing analysis provides 
no evidence for any pulsed X-ray emission from the core of G21.5-0.9.
}
\maketitle

%

\section{Introduction}
\label{sec:intr}

Of the $\sim 225$ known supernova remnants (SNRs) in our galaxy no more
than $5\%$ are classified as Crab-like remnants (Green \cite{green00}). 
Prominent members of this 
class of SNR, also known as plerions, include the Crab Nebula, CTB 87, 
and 3C58. The characteristic centre-filled radio and X-ray morphologies 
of the Crab-like SNRs is thought to be due to the presence of
an active pulsar which powers a bright synchrotron nebula. In the case
of the Crab nebula, 33 ms pulsations testify to the presence 
of the central spinning neutron star, although in many Crab-like systems
there is no direct evidence for pulsed emission.  Typically a Crab-like
SNR exhibits a flat power-law spectrum in the radio regime which eventually 
steepens at shorter wavelengths to join smoothly to a similarly featureless
power-law X-ray continuum. This non-thermal form for the radio to X-ray 
spectrum is a consequence of the continuous injection by the pulsar 
of high energy electrons, which suffer radiation and adiabatic losses
as they diffuse through the nebula; however the details of the process 
are undoubtedly complex (e.g. Reynolds \& Chanan \cite{reynolds84}). 
A feature of the Crab-like remnants, which distinguishes them from both
shell-like and composite-type SNRs (the latter typically exhibiting
a filled-centre X-ray morphology within a radio shell), is the 
lack of any evidence for an outer shell structure marking the progress of 
the blastwave from the original supernova explosion. 

G21.5-0.9 is a SNR with many of the characteristics of
a Crab-like remnant (e.g. Becker \& Szymkowiak \cite{becker81};
Asaoka \& Koyama \cite{asaoka90}). To date there has been no detection 
of pulsed emission in either the radio or X-ray bands
(e.g. Biggs \& Lyne \cite{biggs96}; Slane et al. \cite{slane00}). However, 
recent X-ray observations made by {\em Chandra} have pinpointed
the probable location of the pulsar, namely a very compact 
central core on a scale of $\sim 2''$ embedded within a more extended 
($\sim 30''$ radius) synchrotron nebula (Slane et al. \cite{slane00}). 
A faint extended halo was also detected which might correspond to the 
outer ``shell'' formed from the expanding ejecta and the passage of the 
supernova-driven blastwave (Slane et al. \cite{slane00}). In this 
interpretation G21.5-0.9 would be best described as a composite remnant. 
Here we report further recent X-ray observations of G21.5-0.9 made 
by {\em XMM-Newton} which help to clarify the nature of the extended
X-ray halo in this object.

\begin{figure*}
\centering
\parbox{17.0cm}{\resizebox{\hsize}{!}{\rotatebox{-90}{\includegraphics{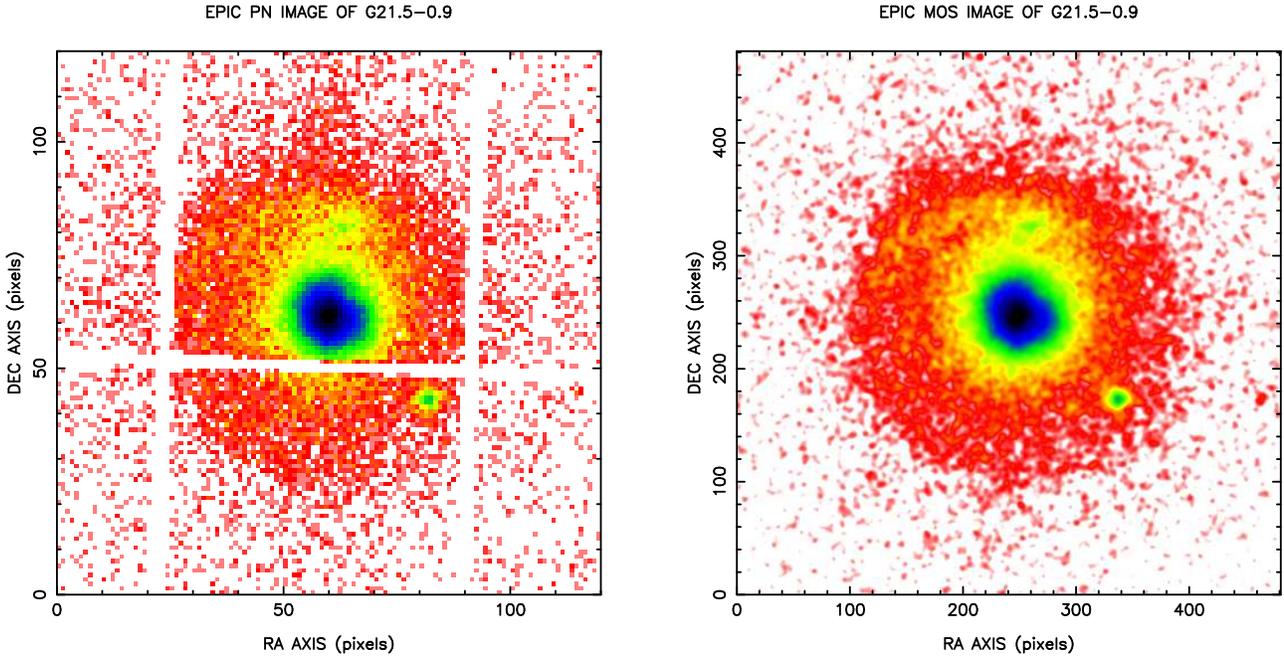}}}}
\hfill
\caption[]{{\it Left panel:} The raw {\em XMM-Newton} EPIC PN image of 
G21.5-0.9. The pixel size is $4''$ and the image is $8'$ on a side. 
A logarithmic intensity scaling has been applied. The vertical and horizontal 
white bands correspond to the gaps between adjacent CCD chips of the PN 
camera. {\it Right panel:} EPIC MOS image of G21.5-0.9. The pixel size 
is $1''$ and the image is $8'$ on a side.  The image has been spatial 
filtered with a Gaussian smoothing mask with width $\sigma = 2 $ pixel.
Again logarithmic intensity scaling has been applied to emphasise 
low surface-brightness features. 
} 
\label{colour}
\end{figure*}

\section{{\em XMM-Newton} observations}

G21.5-0.9 was observed as a calibration target during orbits 60-62 and
64-65 of the {\em XMM-Newton} mission. Here we focus on measurements from 
orbit 60 (date: 2000/04/07; start time: 12.35 UT; end time 22.57 UT) in 
which the SNR was observed on-axis by the EPIC MOS and PN cameras 
(Turner et al. \cite{turner01}; Str{\"u}der et al. \cite{struder01})
giving a total accumulated exposure time of $\sim 30$ ks.  For these 
observations all three EPIC cameras were operated in the standard full-frame 
mode with the medium filter selected.  Attitude information files
are not yet available but all the indications are that a stable pointing was 
maintained for the duration of the observation\footnote{Our
nominal position for the centre of G21.5-0.9 is (RA, Dec) = 18 hr 33 min 
33.8 s, $-10^{\circ}~34'~6''$ (J2000); (l, b) = $21.50^{\circ}$, 
$-0.88^{\circ}$.}.  

The recorded events were 
screened with the XMM Science Analysis Software (SAS) to remove known hot 
pixels and other bad data and pre-processed using the latest CCD gain values. 
X-ray events corresponding to patterns 0--12 for the two MOS cameras (similar 
to grades 0-4 in {\em ASCA}) were used, whereas for the PN only pattern 0 
events (single pixel events) were accepted. Investigation of the full-field 
count-rate revealed a number of flaring events in the non-cosmic component 
of the background during the observation. As a final step in the data 
screening we identified a period when the background was essentially 
``quiescent'', with full-field background rates in the MOS and PN cameras 
of $\sim 2.5$ and $\sim 6 $ count/s.  Selecting events only in this 
low-background period lead to effective exposure times of 13800 s and 
11800 s for the MOS and PN cameras respectively. 


\section{Spatial Analysis}
\label{sec:spa}

Given the observed soft X-ray cut-off in the spectrum of G21.5-0.9 (see \S 4), 
we utilize the 1--10 keV energy band in the spatial analysis. The raw image of 
G21.5-0.9 from the PN dataset binned into $4''$ pixels is shown in Fig. 
\ref{colour} ({\it left panel}). The bright synchrotron nebula at the
centre of this SNR is the most prominent feature of the image, although
the central core is clearly surrounded by a more extended low 
surface-brightness emission region. In addition a point source is evident 
$\sim 2'$ to the south-west of the centre of the remnant.

To avoid the complications due to the chip boundaries in the PN data we have
based our spatial analysis on the data from the MOS cameras (since the 
$10.9' \times 10.9'$ field of view of the central chip in the MOS
cameras comfortably encompasses the full extent of the SNR). Broad band 
images were constructed for both MOS 1 and 2 cameras 
utilizing $1''$ pixels. The MOS 1 and 2 images were then aligned (by eye) 
using the nebula core and the south-western point source as fiducial marks
and co-added. Fig. \ref{colour} ({\it right panel}) shows the resultant image 
after applying a Gaussian smoothing filter with a width $\sigma$ = 2 pixels
(FWHM of $4.7''$). At a qualitative level there is clearly excellent 
agreement between the images from the PN and MOS cameras. The bright centre
of the synchrotron nebula ({\it i.e.} the region within $r \approx 30''$ of 
the centre) shows deviations from circular symmetry, in particular there is
an indentation in the north-west quadrant.  At a slightly lower level of 
surface brightness a curious ``spur'' is evident which appears 
initially to track northwards away from the core region but then at 
$r \sim 100''$ sweeps round in an arc into the north-eastern quadrant of
the remnant. These spatial features match rather well to similar structure
seen in the higher spatial resolution images from Chandra (Slane et al. 
\cite{slane00}). One of the strengths of the EPIC cameras on {\em XMM-Newton} 
is, of course, the sensitivity afforded to extended, relatively low-surface
X-ray emission. This is evident in the images of Fig. \ref{colour} in which
a low surface brightness ``halo'' is well delineated out to 
$r \approx 150''$. This halo exhibits a near perfect circular symmetry apart
from the hint of a ``linear termination'' along its northern rim (note
that out-of-time events, namely events recorded during the period
when charge is transferred to the readout node, partly mask this feature 
in the PN image).

\begin{figure}
\centering
\parbox{7.0cm}{\resizebox{\hsize}{!}{\rotatebox{-90}{\includegraphics{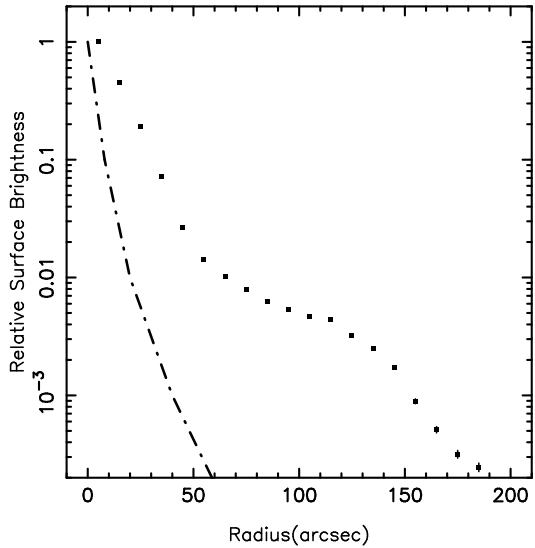}}}}
\hfill
\caption{The radial profile of the X-ray surface brightness of G21.5-0.9 
(dotted curve). The counts in the broad-band MOS image were accumulated 
in annuli of $10''$ width out to a maximum radius of $200''$. The
mean background level in the annular region from $r = 200''$ to $240''$ 
was then subtracted. Note that within a radius of $150''$ 
the statistical errors are smaller than the plotted points. 
The presence of the south-western point source has a marginal impact 
at $r = 115''$. By way of comparison, the point spread function of mirror 
module 3 (MOS 1) at 1.5 keV is also shown (dashed-dotted line) (Aschenbach
et al. \cite{aschenbach00}).
}
\label{profile}
\end{figure}

In order to investigate more quantitatively the spatial extent of the low 
surface brightness halo, we have calculated the azimuthally-averaged radial 
profile of the X-ray emission (Fig. \ref{profile}). The surface brightness 
away from the centre of the synchrotron nebula drops rapidly out to 
$r \approx 50''$. A plateau region is then reached corresponding to the low 
surface brightness halo. A sharp further decline in the surface brightness 
then sets in at $r \approx 130''$, with the signal eventually falling 
below the detection threshold at $r \approx 180''$. In this paper
we define the remnant's outer radius to be $150''$ at which
point the surface brightness is roughly one-fifth of the plateau value.


\section{Spectral Analysis}
\label{sec:spe}

\begin{figure}
\parbox{8.0cm}{\resizebox{\hsize}{!}{\rotatebox{-90}{\includegraphics{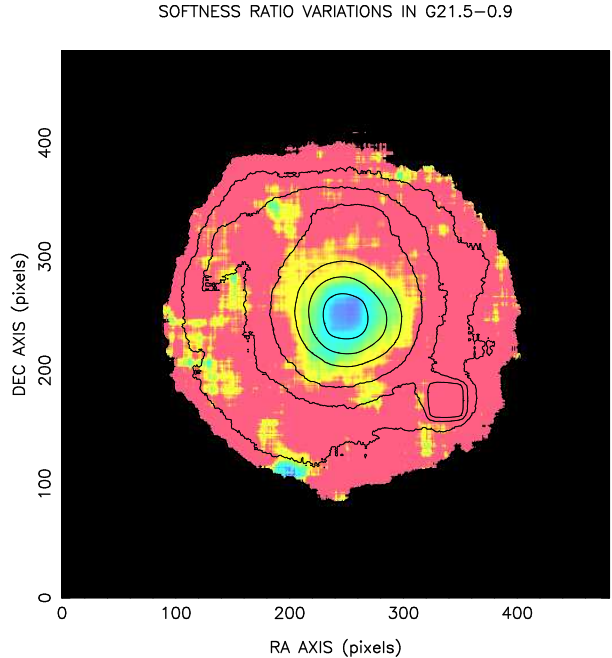}}}}
\hfill
\caption{The softness ratio image of G21.5-0.9 derived from the MOS 1 + MOS 2
datasets. Softness ratio is defined here as SR = (S-H)/(S+H), where S and H 
represent the 1--3 keV and 3--10 keV band images respectively. The S and H 
band images were initially heavily smoothed with a tophat filter of width 
30 pixels (1 pixel = $1''$). Red to blue colours reflected a decreasing
softness ratio. The black contours are based on the combined S+H smoothed 
image.}
\label{softness}
\end{figure}

We first examined the spectral properties of G21.5-0.9 by
plotting the  softness ratio image shown in 
Fig. \ref{softness}, which shows that the core of the nebula is
significantly harder than the low-surface brightness halo. 

We have investigated the spectral characteristics of the core of
G21.5-0.9 using data from both the PN and MOS cameras. The X-ray spectrum of
the core of the nebula was obtained using a circular extraction region
of radius $48''$ centred on the position of peak surface brightness.
Background spectra were extracted from representative regions
greater than $200''$ away from the SNR.  The source spectra were 
binned to a minimum of 20 counts per spectral channel, in order to apply 
$\chi^2$ minimisation techniques. The source and corresponding background
spectra were then analysed using the \textsc{xspec v11.01} and the most 
recent response matrices available from the EPIC team.  The spectral 
modelling assumed a power-law continuum (photon spectral index,
$\Gamma$, and normalisation, $A$) subject to soft X-ray absorption
in cool gas along the line of sight (with an equivalent hydrogen 
column density, $N_{\rm H}$).

\begin{figure}
\centering
\parbox{8.8cm}{\resizebox{\hsize}{!}{\rotatebox{-90}{\includegraphics{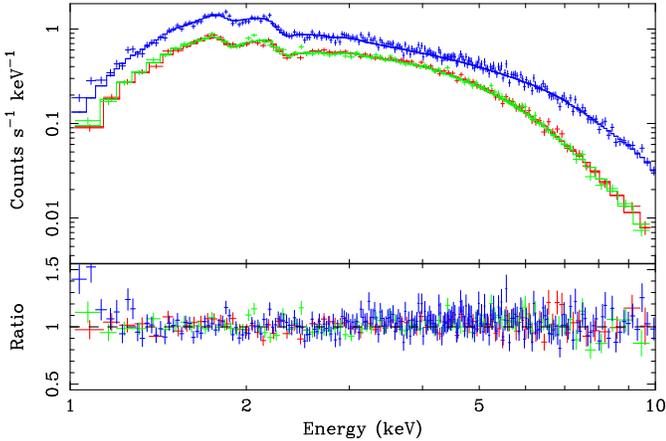}}}}
\hfill
\caption{{\it Upper panel:} The observed PN (blue), MOS 1 (red) and MOS 2 
(green) count rate spectra for the core region of G21.5-0.9. In each case 
the corresponding best fitting spectral model is represented by the solid 
histogram. {\it Lower panel:} The ratio of the predicted to the 
observed count rate spectra.}
\label{spectra}
\end{figure}

Fig. \ref{spectra} shows the observed count rate spectra for the 
PN, MOS 1 and MOS 2 detectors together with the corresponding 
best-fitting model spectra. The spectral response of the two MOS
cameras is so similar that the MOS 1 and MOS 2 count rate spectra
almost totally overlap. The figure illustrates that the observed
count rate in the PN camera is roughly a factor of two 
higher than in a single MOS camera at 2 keV, rising to a factor of four
by 8 keV. The best-fitting spectral parameters obtained for 
separate fits to the PN and MOS spectra are summarised in Table \ref{table_1},
where the errors are quoted at a 90\% confidence level.
The simple absorbed power-law model provides a very good fit to all 
three data sets. In addition the agreement between the spectral parameters
derived independently from the PN, MOS 1 and MOS 2 datasets is excellent.

\begin{table}
\centering   
\caption[]{Best-fitting spectral parameters for the SNR core.}
\begin{minipage}{80 mm}
\begin{tabular}{lcccc}
\hline
\noalign{\smallskip}
Camera  & $N_{\rm H}$$^a$ & $\Gamma$$^b$  & $A$$^c$ & $\chi^{2}$/dof \\
        & & & \\
\noalign{\smallskip}
\hline
\noalign{\smallskip}
 PN    & $2.33^{+0.05}_{-0.05} $ &  $1.86^{+0.03}_{-0.03} $ & $1.56^{+0.08}_{-0.07}$  & 1278/1203 \\
 MOS 1 & $2.28^{+0.06}_{-0.06} $ &  $1.86^{+0.04}_{-0.04} $ & $1.55^{+0.09}_{-0.09}$  & 384/425 \\
 MOS 2 & $2.21^{+0.06}_{-0.06} $ &  $1.86^{+0.04}_{-0.04} $ & $1.61^{+0.09}_{-0.09}$  & 448/423 \\
\noalign{\smallskip}
\hline
\end{tabular}
$^a$ Column density in units of $10^{22}\rm~atom~cm^{-2}$ \\
$^b$ Photon spectral index \\
$^c$ Normalisation in units of $10^{-2}\rm~photon~cm^{-2}~s^{-1}~keV^{-1}$ \\ 
\label{table_1}
\end{minipage}
\end{table}

\begin{figure}
\parbox{8.8cm}{\resizebox{\hsize}{!}{\rotatebox{-90}{\includegraphics{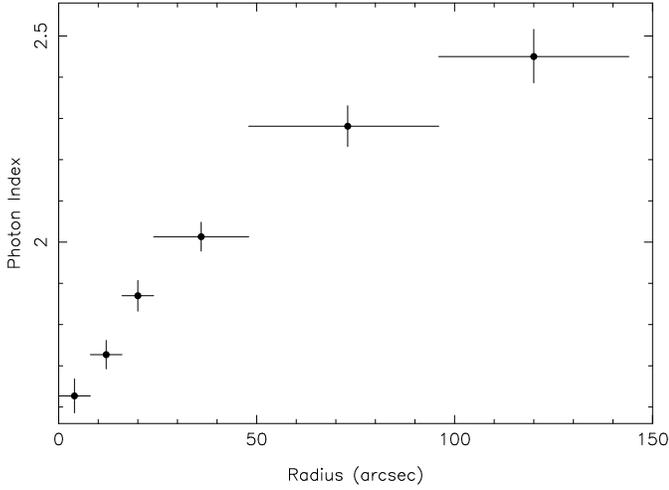}}}}
\hfill
\caption{The observed variation in the photon spectral index versus
radius as measured in the PN data.}
\label{variation}
\end{figure}

\begin{figure}
\centering
\parbox{8.8cm}{\resizebox{\hsize}{!}{\rotatebox{-90}{\includegraphics{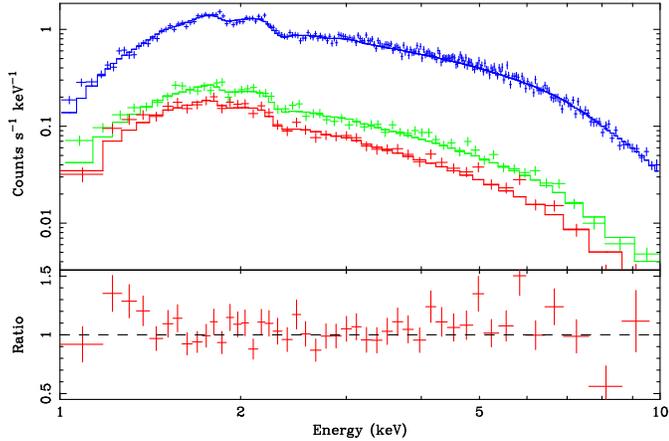}}}}
\hfill
\caption{{\it Upper panel:} The observed count rate spectra and best-fitting 
absorbed power-law model for the core ($r \le 48''$), inner  
($48'' < r \le 96''$) halo and outer ($96'' < r \le 144''$) halo regions 
(shown in blue, green and red respectively). {\it Lower panel:} 
The ratio of the predicted to the observed count rate spectra for the 
outer halo region.}
\label{halo_spec}
\end{figure}

The spectral variations revealed in Fig. \ref{softness} were investigated
by extracting spectra from the PN dataset from a set of annular regions 
centred on the point of peak surface brightness\footnote{In this analysis
we excluded a $16''$ radius region centred on the south-western point 
source.}. The resulting spectra were then fitted as before with a simple
absorbed power-law model. Initially we allowed $N_{\rm H}$ to
vary as a free parameter between the different spectral datasets
but, in the event, excellent fits were obtained with $N_{\rm H}$ fixed
at the value obtained for the core region (Table \ref{table_1}).
The derived photon spectral index shows a steady increase with
radius from a value of $1.63 \pm 0.04$ at the centre to $2.45 \pm 0.06$ 
at the outer edge of the low surface brightness halo (Fig. \ref{variation}).
We note that the rate of spectral softening with radius will be 
slightly underestimated due to spreading caused by the instrument PSF 
(see Fig.\ref{profile}); however the impact on the derived spectral 
indices is likely to be small ($\Delta\Gamma <0.1$). Fig. \ref{halo_spec}
further illustrates the variation of the X-ray spectrum from the core
through to the outer halo region.

We next investigated whether there is any evidence for thermal emission 
from the halo component, focussing on the PN spectrum extracted from an 
outer ($r = 96'' - 144''$) annular region. As a first step we fitted
a two component model, comprising  a power-law and a (solar abundance, MEKAL) 
thermal component. The result was a modest improvement in the $\chi^{2}$ with 
respect to the power-law model (see Table \ref{table_2}) 
which the F-test shows to be marginally significant for two additional 
parameters. However, this putative ($\sim 1$ keV) thermal component 
accounts for only $\sim 1\%$ of the observed flux. 
The lack of prominent line features in the X-ray spectrum of the outer halo
is evident in Fig. \ref{halo_spec} which shows the fitting residuals
to a simple power-law spectral model. One possibility is that the line 
emission is suppressed  due to the fact that shock-heated X-ray emitting gas 
in the halo of G21.5-0.9 has not had time to reach a state of ionization 
equilibrium. When we fit a non-equilibrium ionization model (the NEI 
model in \textsc{xspec}) we obtain the spectral parameters listed in Table 
\ref{table_2}. The NEI model provides only a rather poor fit to the data
when the column density is constrained to the value measured for the SNR core.
A rather better result is obtained when $N_{\rm H}$ is allowed to vary freely  
(the last entry in Table \ref{table_2}) but nevertheless the fit is not
as good as that obtained for the absorbed power-law model.

We also examined the spectrum of the brightest part of the spur feature 
(located directly to the north of the core) but unfortunately the 
signal-to-noise ratio was too low to distinguish between thermal and 
power-law spectral forms.

\begin{table} 
\centering
\begin{minipage}{88 mm}
\caption[]{Best-fitting spectral parameters for the outer halo region
of G21.5-0.9}
\begin{tabular}{lccccc}
\hline
\noalign{\smallskip}
Model   & $N_{\rm H}$$^a$ & $\Gamma$$^b$ & $kT$$^c$ & $n_{\rm e}t$$^d$ & $\chi^{2}$/dof \\
        &           &              &          &            &                \\
\noalign{\smallskip}
\hline
\noalign{\smallskip}
PL& $2.33$$^e$ & $2.45^{+0.06}_{-0.06}$  & - & - &  258/231  \\
2-Comp & $2.33$$^e$ & $2.39^{+0.12}_{-0.12}$  & $\sim 1 $ & - & 253/229 \\
NEI & $2.33$$^e$ &  -   & $3.1^{+0.2}_{-0.2}$ & $7^{+3}_{-3}$  & 356/230  \\
NEI & $1.69^{+0.11}_{-0.11}$  & - & $4.4^{+0.5}_{-0.5}$ & $3^{+2}_{-3}$ & 277/229  \\
\noalign{\smallskip}
\hline
\end{tabular}
$^a$ Column density in units of $10^{22}\rm~atom~cm^{-2}$ \\
$^b$ Photon spectral index \\
$^c$ Temperature in keV \\
$^d$ Ionization timescale in units of $10^{8} \rm~cm^{-3}~s$ \\
$^e$ Fixed parameter
\label{table_2}
\end{minipage}
\end{table}

The total integrated X-ray flux of G21.5-0.9 within a radius of $144''$,
corrected for the line-of-sight absorption, is $6.1 \  \times 
10^{-11}$~erg~cm$^{-2} $~s$^{-1}$ in the 2--10 keV band. 
Assuming a distance of 5 kpc based on HI absorption measurements 
(Davelaar, Smith \& Becker \cite{davelaar86}), this implies an
X-ray luminosity of $1.8 \times 10^{35} \rm~erg~s^{-1}$.

Finally we have briefly considered the spectral properties of the
south-west point source. We obtained an on-source spectrum from 
the PN dataset using an extraction cell of  $16''$ radius. A 
complementary background spectrum was taken (on the same CCD) 
using a cell of the same dimension sited equidistant from the centre 
of G21.5-0.9.
Spectral fitting indicated a relatively hard spectrum ($ \Gamma = 
1.61\pm 0.34$ or $\rm kT \sim 7$ keV) and a column density of  $1.03\pm 0.22
\times 10^{22} \rm~cm^{-2}$. As the latter is about a factor
two lower than the derived $N_{\rm H}$ of G21.5-0.9, we can conclude
that this point source is a foreground object with respect to the SNR. 
If we estimated its distance as $\sim 3$ kpc ({\it i.e.} roughly half the 
distance to the SNR in line with the ratio of $N_{\rm H}$ for the two sources) 
then the inferred 2--10 keV luminosity is 
$4 \times 10^{32} \rm~erg~s^{-1}$. This point source is positionally 
coincident with the emission-line star SS397.


\section{X-ray Timing Analysis}

We have carried out a preliminary search for pulsed X-ray emission 
using EPIC data from the {\it full} observation  ({\it i.e.} including 
periods of relatively high background rate). The photon events in the 
1--10 keV band from a region of $8''$ radius centred on the core 
of G21.5-0.9 were extracted for all three cameras. Next the event times were 
barycentric corrected using the standard SAS task ``Barycen'' and the 
latest available ``Reconstructed Orbit'' file. Power spectra were then 
calculated for the combined data from all three cameras and the PN dataset
alone. No evidence was found for a significant periodic signal. In terms of 
the amplitude of an underlying sinusoidal signal, the detection threshold 
was $\sim 5 \times 10^{-2}$ ct/s in the frequency range 
$5 \times 10^{-4}$ -- 0.17 Hz for the PN + MOS  combination 
(the upper frequency limit being set by the frame time of the MOS CCDs). 
Corresponding values for the PN-only data were $\sim 3 \times 
10^{-2}$ ct/s in the frequency range $5 \times 10^{-4}$ -- 2 Hz. 
In terms of the pulsed fraction (of the core emission)
the limits are roughly $3.5\%$ and $5.5\%$ respectively.


\section{Discussion}

In a recent paper reporting Chandra observations, Slane et al.
(\cite{slane00}) consider the possibility that the low-surface brightness 
halo surrounding the central synchrotron nebula in G21.5-0.9 might
be the shell formed from the ejecta and blastwave of
the original supernova explosion.  The present {\em XMM-Newton} observations 
help delineate the spatial extent and morphology of this component.
Crucially the new spectral information demonstrates that the extended halo 
has a spectral form devoid of any significant line features. If the halo 
emission is thermal then the lack of line emission implies that the plasma 
is far from ionization equilibrium. Our NEI modelling confirms
a low ionization state with $n_{\rm e}t \approx 3 \times 10^{8} \rm~cm^{-3}~s$.
The temperature derived for the continuum (bremsstrahlung) emission is
$4 - 5$ keV which is rather hot, even for a very young SNR. Assuming the 
distance to 
G21.5-0.9 is 5 kpc (Davelaar, Smith \& Becker \cite{davelaar86}) and taking 
a maximum shell expansion velocity of $10,000 \rm~km~s^{-1}$, 
we can estimate the (minimum) elapse time since the shock heating of the 
bulk of the X-ray emitting gas in the outer halo to be $\sim 100$ yrs. 
Applying the above constraint on $n_{\rm e}t$ then gives
$n_{\rm e} < 0.1 \rm~cm^{-3}$. However, the observed X-ray luminosity of 
the outer halo, if interpreted as bremsstrallung radiation, sets a 
requirement for a significantly higher electron density, 
$n_{\rm e} \approx 0.5 \rm~cm^{-3}$. There would appear therefore to be
considerable difficulties in interpreting the extended X-ray halo in 
G21.5-0.9 as a thermal shell.

There are known examples of SNR shells which exhibit non-thermal 
X-ray spectra due to local particle acceleration (e.g. SN 1006, 
Koyama et al. \cite{koyama95}; G347.3-0.5, Slane et al. \cite{slane99}). 
However, in the case of G21.5-0.9 the lack of any limb brightening, the 
smooth transition in spectral index throughout the remnant and the 
remarkable circular symmetry, all suggest an interpretation of the outer halo 
as an extension of the central synchrotron nebula. We conclude therefore, 
on the basis of the present observations, that G21.5-0.9 is a true 
Crab-like system rather than a composite (centre-filled plus thermal shell) 
object.

The observed spectral softening with radius in G21.5-0.9 most 
likely reflects the impact of synchrotron radiation losses on very high 
energy electrons as they flow from the region of the termination shock 
around the pulsar to the edges of the nebula. For example, if the nebula 
magnetic field is  $\sim 0.4 $ mG (Slane et al. \cite{slane00} ), then the 
synchrotron lifetime of the $\sim 20$ TeV electrons producing 5 keV X-rays 
is only $\sim 2$ yrs. Similar spectral trends due to ``synchrotron burn-off''
are seen in other Crab-like SNR, such as 3C58 (Torii et al. \cite{torii00}). 
In fact the variation in photon spectral index from a value of 1.63 at the 
centre of the G21.5-0.9 to 2.45 at the edge of its halo is amazingly 
similar to the index range seen in the Crab nebula (Willingale et al. 
\cite{willingale01}). In the case of the Crab nebula the spatial and spectral 
distribution of the optical to hard X-ray continuum has been explained in 
terms a magnetohydrodynamic flow of plasma and magnetic flux in a particle 
dominated wind (e.g. Kennel \& Coroniti \cite{kennel84}). For G21.5-0.9, 
the origin of halo which represents a ``plateau'' in the surface 
brightness extending from $r \approx 50'' - 130''$ remains unclear, 
but could possibly be the result of an unusual magnetic field geometry 
(see below).

One interesting contrast is that the radio extent of the Crab nebula
is roughly a factor four greater than that observed in X-rays, whereas for 
G21.5-0.9 the dimensions of the X-ray halo exceeds that of the radio 
synchrotron nebula (which has a major axis of $\sim 90''$;  
F{\"u}rst et al. \cite{furst88}) by a similar factor. Slane et al. 
(\cite{slane00}) note that no radio ``shell'' is detectable to a (1$\sigma$)
threshold of $4 \times 10^{-21} \rm~W~m^{-2}~Hz^{-2}~sr^{-1}$ at 1 GHz. 
If we make the assumption that the ratio of the core to halo flux is the 
same in the X-ray and radio regimes, then we predict the average
surface brightness of an underlying radio halo (between $r = 60'' - 130 ''$) 
to be roughly 5 times this threshold.  The inference is that very deep 
radio observations may well reveal the presence of an extended radio halo
in G21.5-0.9. 

The differences in the spatial morphology of the Crab 
and G21.5-0.9 could be due to the viewing orientation. Specifically the near 
circular symmetry of G21.5-0.9 might be explained if our line of sight is
reasonably aligned with the spin axis of the resident pulsar, so as to give 
a near face-on view of a putative inner torus.  This hypothesis might also 
explain the non-detection of pulsed X-ray emission from the core of G21.5-0.9,
but is hard to reconcile with the axisymmetric structure seen in high 
frequency radio maps and the near radial distribution of the magnetic field 
lines (F{\"u}rst et al. \cite{furst88}). In this setting the spur-like 
feature which emanates northwards from the core could a ``plerionic'' wisp 
associated with a complex inner magnetic field and/or jet structure.





\begin{acknowledgements}

The results presented in this paper are based on observations obtained 
with XMM-Newton, an ESA science mission with instruments and contributions 
directly funded by ESA Member States and the USA (NASA). We thank the 
whole {\em XMM-Newton team} for the hard work and dedication which 
underlies the success of the mission.

\end{acknowledgements}


\end{document}